\documentstyle[12pt]{article}
\begin{document}
\begin{center}
{\Large \bf Can QFT be UV finite as well as effective?
}
\bigskip
{\large

Jifeng ~Yang

}
\date{}
\smallskip
{\it
School of Management, Fudan University,
Shanghai 200433, P. R. China  }
\end{center}
\begin{abstract}
A new attempt is demonstrated that QFTs can be UV finite if
they are viewed as the low energy effective theories of a
fundamental underlying theory (complete and well-defined
in all respects) according to the modern standard point of
view. This approach is much simpler in principle and in
technology comparing to any known renormalization program.
Some subtle and difficult issues can be easily resolved.
The importance of the procedure for defining the
ambiguities is fully appreciated in the new approach. Some simple
but important nonperturbative examples are discussed to show
the power and plausibility of the new approach.
\end{abstract}
\vspace{1.5cm}

Now, it has become a standard point of view that a
fundamental theory (well defined in every aspect) underlies
the present QFTs that are in fact low energy (LE) effective
theories for the phenomena in LE ranges \cite {Jac}. But as
far as the author knows, we are still lacking a formulation
that can yield finite results in a natural way making use of
this point of view. Here I would like to present a new
approach directly based on this principle \cite {YYY}.

In this point of view, all the Feynman amplitudes (FAs) or
the various Green
functions given by a present QFT should correspond to
the LE limits of certain subset of the well-defined
'Green functions' given by the underlying theory.
Employing a generating functional or path integral
formalism \cite {Shirkov} to assemble these Green
functions for each of the subsets, the true formulation
should be
\begin{eqnarray}
Z^0 (\{J^i\};\overline{\cdots})
&\equiv& {\bf L}_{\{\sigma\}}
   \left \{ \int D{\mu}(\phi^{i}_{\{\sigma\}})
    \exp \{iS(\phi^{i}_{\{\sigma\}};\{J_{i}\};\{\sigma\})\}
    \right \}  \\
&\not=&\int D{\mu}(\phi^i)\exp\{iS(\phi^i;\{J_i\})\},
\end{eqnarray}
where $\{\sigma\}$ are the underlying fundamental parameters
in the underlying theory and $\{J^{i}\}$ are the external
sources specifying the LE phenomenon. The 'elementary fields'
and the action for the QFTs are the LE limits of the
corresponding ones derived from the underlying theory that
depend on the underlying parameters. The dots with an overline
represents the possible remaining effects of the underlying
parameters($\{\sigma\}$). The inequality (2) is just the origin
of UV ill-definedness, {\it the non-commutativity of the LE limit
operation and the summation over the 'paths'}. Only the
convergent FAs are indifferent to the order of operations. The
present QFTs are just ill-defined LE {\it reformulations or
reorganizations} of these subsets of FAs or Green functions.

In the following I will sketch a 'recipe' for getting rid of the 
illdefinedness naturally following from the standard point of 
view with the existence of the underlying theory that is well 
defined in every aspect.

First we show that the following important relation holds for
1-loop ill-defined FAs (1-loop divergent Feynman graphs)
\begin{equation}
\int d^{n}Q \left ({\partial}_{p_{j}} \right )^{\omega}
f(Q,\{p_{j}\},\{m_{k}\})=
\left ( {\partial}_{p_{j}} \right )^{\omega} \Gamma^{0}
(\{p_{j}\},\{m_{k}\}),
\end{equation}
with $\omega-1$ being the usual superficial divergence
degree of $\int d^{n}Q f (Q,\{p_{j}\},\{m_{k}\})$ so that the
lhs of Eq.(3) exists (finite),
$\left ({\partial}_{p_{j}} \right )^{\omega} $
denoting differentiation's wrt the external parameters
$\{p_{j}\}$'s of the amplitude and $\Gamma^{0}(...)$ is the
LE limit of the amplitude calculated in the underlying theory
(i.e., the internal momentum integration is performed first).

The proof is very simple, since
\begin{eqnarray}
&\int& d^{n}Q  \left ({\partial}_{p_{j}} \right )^{\omega}
f (Q,\{p_{j}\},\{m_{k}\})= \int d^{n}Q
\left ({\partial}_{p_{j}} \right )^{\omega} {\bf  L}_{\{\sigma\}}
\bar{f} (Q,\{p_{j}\},\{m_{k}\};\{\sigma_{l}\})\nonumber \\
&=& \int d^{n}Q {\bf  L}_{\{\sigma\}} \left ({\partial}_{p_{j}}
\right )^{\omega}\bar{f}
(Q,\{p_{j}\},\{m_{k}\};\{\sigma_{l}\})\nonumber \\
&=& {\bf  L}_{\{\sigma\}}  \int d^{n}Q
\left ({\partial}_{p_{j}} \right )^{\omega}\bar{f}
(Q,\{p_{j}\},\{m_{k}\};\{\sigma_{l}\})\nonumber \\
&=&{\bf L}_{\{\sigma\}} \left ({\partial}_{p_{j}} \right )^{\omega}
\overline{\Gamma} (\{p_{j}\},\{m_{k}\};\{\sigma_{l}\}) =
\left ({\partial}_{p_{j}} \right )^{\omega} \Gamma^{0}
 (\{p_{j}\},\{m_{k}\}).
\end{eqnarray}
The second and the fifth steps follow from the commutativity of
the two operations $\left ({\partial}_{p_{j}} \right )^{\omega}$
and ${\bf  L}_{\{\sigma\}}$ as they act on different arguments,
the third step is due to the existence of
$\int d^{n}Q \left ({\partial}_{p_{j}}
\right )^{\omega} f (Q,...)$ and the fourth is justified from
the existence of $\int d^{n}Q \bar{f} (Q,...;\{\sigma_{l}\})
( = \overline {\Gamma} (...;\{\sigma_{l}\}))$.

The right end of Eq.(3) can be found now as the left end exists
as a nonpolynomial (nonlocal) function of external momenta
and masses. To find $\Gamma^{0} (\{p_{j}\},\{m_{k}\})$, we
integrate both sides of Eq.(3) wrt the external momenta
"$\omega$" times indefinitely to arrive at the following
expressions
\begin{eqnarray}
& &\left (\int_{{p}}\right )^{\omega}
 \left [ ({\partial}_{{p}})^{\omega} \Gamma^{0}
(\{p_{j}\},\{m_{k}\}) \right ] = \Gamma^{0} (\{p_{j}\},\{m_{k}\})
 +  N^{\omega} (\{p_{j}\},\{c_{\omega}\}) \nonumber \\
&=& \Gamma_{npl} (\{p_{j}\},\{m_{k}\}) + N^{\omega}
(\{p_{j}\}, \{C_{\omega}\})
\end{eqnarray}
with $\{c_{\omega}\}$ and $\{C_{\omega}\}$ being arbitrary
constant coefficients of an $\omega-1$ order polynomial in
external momenta $N^{\omega}$ and $\Gamma_{npl}$ being a definite
nonpolynomial function of momenta and masses. Evidently
$\Gamma^{0}$ is not uniquely obtained within conventional QFTs
at this stage, its true expression should contain a definite
polynomial part ($N^{\omega}(\cdots;\{\bar{c}_{\omega}\})$ with
$\bar{c}_{\omega}=C_{\omega}-c_{\omega}$) (which is unknown yet
and thus blurred by ambiguities ($\{c_{\omega}\}$)) implying that
it should have come from the LE limit operation.

We can take the above procedures as efforts for rectifying
the ill-defined FAs, i.e.,
\begin{equation}
\int d^{n}Q f (Q,\{p_{j}\},\{m_{k}\}) >=< \Gamma_{npl}
(\{p_{j}\},\{m_{k}\}) + N^{\omega} (\{p_{j}\}, \{C_{\omega}\})
\end{equation}
with "$>=<$" indicating that lhs is rectified as rhs \cite {JF}.
Although ambiguous, the rhs of Eq.(6) is the best we can achieve
within the present QFTs.

As we have addressed, the $\bar{c}_{\omega} $'s arise in fact
from the LE limit operation, they should be uniquely defined for
any specific low energy phenomenology up to possible equivalence.
Different (inequivalent) choices of these constants either are
incorrect or simply lead to irrelevant predictions. Since
different Regs and/or Ren conditions just correspond to
different choices of the constants, we may find, especially
in nonperturbative cases (such important examples can be found
in Ref. \cite {QMDR}), that they would lead to rather different
'renormalized' LE theories, or even could not describe relevant
low energy physics. Thus, {\it the underlying theory does 'stipulate'
or influence the effective ones through these constants though the
underlying parameters do not explicitly appear in the LE formulations.}
All the known programs failed to fully appreciate this important issue.
For further discussions, see Ref. \cite {YYY,Talk3}.

Now we consider the treatment for the multi-loop case which is
very simple and straightforward.

Suppose a multi-loop graph $\Gamma$ (we will
use the same symbol to denote the graph and the FA
associated with it if it is not confusing) contains at least
overall divergence, we proceed like the following (just like
in the 1-loop case), (we will use in the following
$\omega_{\gamma}-1$ to denote the overall divergence index
\cite {Shirkov} for any graph $\gamma$ and $\{l\}$ to refer to
the internal momenta, all the partial differentiation
operators and their 'inverse' (denoted by
$\partial_{\omega_\gamma}^{-1}$) act upon the
momenta only external to the very internal integration
of the graph under consideration)
\begin{eqnarray}
& &\Gamma^0(\cdots;\{c^0_{i}\})\equiv{\bf L}_{\{\sigma\}} \int
   \prod dl \bar{f}_{\Gamma}(\{l\},\cdots;\{\sigma\})
    \hfill \nonumber \\
&\Rightarrow&\partial_{\omega_\Gamma}^{-1} \{
    {\bf L}_{\{\sigma\}} \int \prod dl \partial^{\omega_{\Gamma}}
    \bar{f}_{\Gamma} (\{l\},\cdots; \{\sigma\}) \} \\
&=&\sum_{\{\gamma\}=\partial^{\omega_{\Gamma}}\Gamma}
   \partial_{\omega_\Gamma}^{-1} \left \{ {\bf L}_{\{\sigma\}} \int
   \prod dl \bar{f}_{\gamma} (\{l\}, \cdots; \{\sigma\}) \right \}\\
&=&\sum_{\{\gamma\}}\partial_{\omega_\Gamma}^{-1} \{\int
   \prod d \overline {l^{\prime}} g_{\gamma/{[\gamma^{\prime} ]}}
   (\{\overline{l^{\prime}}\}, \cdots)
   {\bf L}_{\{\sigma\}} [ \prod_{\gamma^{\prime}_{j}}
    \int \prod_{i\epsilon \gamma^{\prime}_{j}}
    dl^{\prime}_{i} {\bar {f}}_{\gamma: \gamma^{\prime}_{j}}
    (\{l^{\prime}\}_{\gamma^{\prime}_{j}},\cdots;
    \{\sigma\}) ] \} \nonumber \\
&=&\sum_{\{\gamma\}}\partial_{\omega_\Gamma}^{-1} \{ \int \prod d
    \overline{l^{\prime}}g_{\gamma/{[\gamma^{\prime}]}}
   (\{\overline{l^{\prime}}\}, \cdots)
   \prod_{\gamma^{\prime}_{j}} [ {\bf L}_{\{\sigma\}}
   {\overline{\Gamma}}_{\gamma: \gamma^{\prime}_{j}}
    (\cdots;\{\sigma\}) ] \}, \\
& &(\bigcup_{\gamma^{\prime}_{j}}
   \{l^{\prime}\}_{\gamma^{\prime}_{j}})
   \bigcup \{ \overline{l^{\prime}}\}=\{l\},\ [\gamma^{\prime}]
   \bigcup (\gamma/{[\gamma^{\prime}]} )=\gamma, \
   [\gamma^{\prime}]=\prod_{j} \gamma^{\prime}_{j},\nonumber \\
& &\gamma^{\prime}_{j}\bigcap \gamma^{\prime}_{k} =0 \ \
   (j \not= k).
\end{eqnarray}
Here we note that the differentiation wrt the external parameters
'created' a sum of graphs $\{\gamma\}$ (without overall divergence)
from the original graph $\Gamma$. Any overall divergence (including
overally overlapping ones) is hence killed \cite {CK}. Each graph in
the set ${\partial}^{\omega_{\Gamma}}\Gamma$ is a 'product' of
disconnected subgraphs (each subgraph itself may contain overlapping
divergences). The LE limit operator crossed all the other parts but
stopped before the divergent subgraphs. All the dots in the
expressions refer to the parameters (some of them are themselves
internal momenta to some (sub)graphs) 'external' to the loop
integrations for the subgraphs. Since some loop momenta are
'external' to other subgraphs, {\it one should not carry out these
loop integrations before the ill-defined subgraphs are treated.}

As the ill-defined subgraphs in $[\gamma^{\prime}]$ are
disconnected with each other, we now treat each of them
separately as a new 'total' graph and go through the
procedures from Eq.(7) to Eq.(10) again and again till we
meet the smallest subgraphs that are completely convergent.
In this course, the LE limit operator crosses more and more
parts of the total graph till the final part, then we get
the integrands totally expressed with propagators and
vertices given by the effective theories and we can begin
to perform all the loop integrations backwardly, i.e.,
from the smallest subgraphs up to larger graphs till all the
internal loop integrations and all the 'inverse' operations
(indefinite integration) wrt various 'external' momenta
({\it after} the associated internal integrations) are done, a
{\it natural} order from our treatment. [It is worthwhile to note
that at each level of the subgraphs, the loop integrations are
guaranteed to be convergent due to Weinberg's theorem
\cite {weinth}].

The resulting expression will be a definite nonlocal function
plus nonlocal ambiguities (due to subgraph divergences) and
local ambiguities if $\Gamma$ is suffering from overall
divergences.

Here some remarks are in order.

{\bf A}. It is evident that overlapping divergences are
just automatically resolved in our approach, there is
nothing special about it.

{\bf B}. In our treatment of the ill-defined graphs, since
every loop integration actually performed is convergent,
the linear transformations of the internal momenta do not
alter the results of the loop integrations. Due to the
'inverse' operator, these linear transformations of the
integration variables will at most change the ambiguous
constants that are yet to be determined. Thus these linear
transformations merely lead to Reg effects.
This observation implies that one should not worry about the
variable shifting and routing of the external momenta that
belong to the transformations just described. For further
physical implications of this property see Ref. \cite {Talk3}.

Here I would like to cite a recent work that fully
exhibits the difficulties with the old Reg and Ren schemes,
the work by Phillips {\it et al} \cite {QMDR} where it is
demonstrated that in the nonperturbative contexts the cutoff
scheme and the dimensional regularization (DR) scheme are
inequivalent to each other. This is due to that different
Reg schemes yield different degrees of divergences which in
turn 'deform' the theory quite differently (inequivalent
choices of the constants). More irremovable divergences in
a Reg scheme means less rationality in using it, and hence
the cutoff scheme has no better chance than the DR scheme
to yield reasonable conclusions.

Reexamining the problem within our approach one would revive
the 'effective field theory' approach to the LE nuclear
physics initiated by Weinberg \cite {WeinEFT} from the recent
criticism \cite {Cohen}(for more details and discussions, see
\cite {Talk3}).

As a matter of fact the QM Hamiltonian with singular
potentials (like the Delta-potentials) needs self-adjoint
extension and an additional parameter--the family
parameter--necessarily appears which upon different choices
leads to different or inequivalent (LE) physics \cite {QM},
just supporting our point of view. Thus, such nonperturbative
problems can be critical touchstones for these schemes.

For the IR problem please refer to Ref. \cite {Talk3} where
a new possibility of treatment is suggested basing on a similar
idea that the present QFTs go wrong in the IR ends of the spectra.

In summary, we discussed briefly the approach
recently proposed by the author and the important consequences
following from it. We have overcome many typical difficulties and
shortcomings associated with old Reg and Ren frameworks. The
method is simple and powerful in many respects.

\end{document}